\documentstyle{article}

\begin{document}

\title{Global Linear Complexity Analysis of Filter Keystream Generators}
\date{}
\author{ A. F\'uster-Sabater(1) and P. Caballero-Gil(2)\\
{\small(1)Consejo Superior de Investigaciones Cient\'ificas, Madrid, Spain} \\
{\small(2)Universidad de La Laguna, Facultad de Matem\'aticas, }\\
{\small Departamento de Estad\'istica, IO y Computaci\'on, La Laguna, Spain}}
\maketitle

\begin{abstract}
An efficient algorithm for computing lower bounds on the global linear 
complexity
of nonlinearly filtered PN-sequences is presented. 
The technique here developed is based exclusively on the realization of bit
wise logic operations, which makes it appropriate for both software 
simulation
and hardware implementation. The present algorithm can be applied to any 
arbitrary nonlinear function with a unique term of maximum order.
Thus, the extent of its application for different types of filter generators 
is quite broad.
Furthermore, 
emphasis is on the large lower bounds obtained that confirm the exponential
growth of the global linear complexity for the class of nonlinearly filtered
sequences.
\end{abstract}

\section{Introduction}
\footnotetext{
This work was supported by R\&D Spanish Program TIC95-0080.\\
IEE Proceedings Computers and Digital Techniques, January 1997, Vol. 144, Is. 1, p.33-38.\\ 
DOI: 10.1049/ip-cdt:19970764 
}

Many procedures in modern communication systems require binary sequences which
appear to be random but, in fact, have been generated in a deterministic way.
They are the so-called pseudorandom sequences. In cryptographic applications 
the sequence obtained in such a way is referred to as the keystream.
To provide secure encryption the keystream must verify several properties
of cryptographic nature such as: long periods, balanced statistics, mth-order 
correlation immunity, distance to linear functions, avalanche criterion... 
(for a more detailed survey see \cite{Simmons}).
In addition a keystream generator has to be unpredictable: that is, given
a portion of the output sequence, a cryptoanalyst should be unable to 
predict other bits forward or backward. A widely accepted measure of the 
unpredictability
of a sequence is the linear complexity defined as the shortest linear 
recursion over GF(2) satisfied by such a sequence.

One of the most commonly used  keystream generators  is obtained by
applying a nonlinear function to the stages of a maximal-length Linear 
Feedback 
Shift Register (LFSR). 
This type of generator is called `filter generator'. The linear complexity of 
the resulting keystream can be computed in two
different ways:

1.- Analysing the digits of the output sequence by means of the
Berlekamp-Massey LFSR synthesis algorithm \cite{Massey}.

2.- Studying the nonlinear function applied to the LFSR's stages.

Local linear complexity and global linear complexity are obtained in each case 
respectively. The global linear complexity of the filter generators depends 
exclusively on the particular form of the filter and the
LFSR's minimal polynomial. Generally speaking, there is no systematic method
to predict the resulting global linear complexity. This is the reason why in
the open literature statements like `it is extremely difficult to lowerbound
(or guarantee) the linear complexity of the sequences produced by
nonlinearly filtering the state of an LFSR' \cite[pp. 57]{Rueppel} can be
found. Nevertheless, some authors have faced this problem and several
references can be quoted. Apart from the works of Groth \cite{Groth} and 
Key \cite{Key}, Kumar
and Scholtz \cite{Kumar-Scholtz}
derived a general lower bound for the class of 
bent sequences, although the LFSR's length is restricted to be a multiple
of 4.
Rueppel \cite{Rueppel} established his 
{\it root presence test} for the product of distinct phases of a
PN-sequence, which is based on the computation of determinants in a finite
field. One of the most recent works on this subject, \cite{Massey-Serconek},
has focussed on the use of the Discrete Fourier Transform Technique to
analyse the global linear complexity. Most of the above mentioned works impose
rather restrictive conditions on the LFSR's length, the order of the nonlinear
function or the particular form of the applied function.

Based on the works \cite{Rueppel} and \cite
{Fuster-Caballero}, a new algorithm (the so-called LB-algorithm) is
proposed for the computation of  lower bounds on the global linear
complexity. This algorithm can be applied to any arbitrary 
nonlinear filter with a
unique term of maximum order. In fact, no restrictions are imposed on the 
LFSR's  stages, the particular form of the filter or the LFSR's minimal
polynomial. On the other hand, the most important feature of the 
LB-algorithm is that it  is based exclusively on the realization of bit wise 
logic
operations (OR, AND and XOR), which makes it rather adequate to either
software simulation or hardware implementation.

As the algorithm INPUTS are L (LFSR's length) and k (order of the function),
then the lower bound obtained is valid for any kth-order function with a
unique term of maximum order and for any LFSR of length L.

\section{Review of the Root Presence Test and new Definitions}

Some fundamental concepts and notation which are used in this work can be
introduced as follows.

$S$ is the output sequence of an LFSR whose minimal polynomial $m_s(x)\in
GF(2)[x]$ is primitive. $L$ is the length of the LFSR. $\alpha \in GF(2^L)$
is one root of $m_s(x)$. $f_k$ denotes the unique maximum order term of a
nonlinear kth-order function $f$ applied to the LFSR's stages, $
f_k=s_{n+t_0}s_{n+t_1}\cdot \cdot \cdot s_{n+t_{k-1}}$ where the symbols $
t_j $ (j=0,1,...,k-1) are integers verifying $0\leq t_0<t_1<\cdot \cdot
\cdot <t_{k-1}<2^L-1$. In this work only the contribution of $f_k$ to the
global linear complexity of the resulting sequence will be studied.

The {\it root presence test} for the product of k distinct phases of a
PN-sequence can be stated as follows, \cite{Rueppel}:

$\alpha ^E\in GF(2^L)$ is a root of the minimal polynomial of the generated
sequence if and only if 
$$
A_E=\left| 
\begin{array}{c}
\alpha ^{t_02^{e_0}} \\ 
\alpha ^{t_02^{e_1}} \\ 
. \\ 
\alpha ^{t_02^{e_{k-1}}} 
\end{array}
\begin{array}{c}
\alpha ^{t_12^{e_0}} \\ 
\alpha ^{t_12^{e_1}} \\ 
. \\ 
\alpha ^{t_12^{e_{k-1}}} 
\end{array}
\begin{array}{cccc}
. & . & . & \alpha ^{t_{k-1}2^{e_0}} \\ 
. & . & . & \alpha ^{t_{k-1}2^{e_1}} \\ 
. & . & . & . \\ 
. & . & . & \alpha ^{t_{k-1}2^{e_{k-1}}} 
\end{array}
\right| \neq 0 
$$

Here $\alpha ^{t_j}\in GF(2^L)$ (j=0,1,..,k-1) correspond respectively to
the k phases $(s_{n+t_j})$ of the PN-sequence. {\it E}, the representative
element of the cyclotomic {\it coset E}, is a positive integer of the form $
E=2^{e_0}+2^{e_1}+\cdot \cdot \cdot +2^{e_{k-1}}$ with the $e_i\ $
(i=0,1,..., k-1) all different running in the interval $[0,L).$
Under these conditions, $\alpha ^E$ and its conjugate roots contribute to
the global linear complexity of the nonlinearly filtered sequence. The value 
of
this contribution is equal to the number of elements in such a cyclotomic
coset.

The cyclotomic {\it coset E }is said to be {\it degenerate} 
if the corresponding determinant $A_E$ equals zero. Otherwise the cyclotomic 
{\it coset E} will be {\it nondegenerate}.

Notice that every cyclotomic {\it coset E} can be easily associated with the
radix-2 form of the integer {\it E}. This fact quite naturally suggests the
introduction of binary strings of length L and Hamming weight k. Indeed, the
cyclotomic {\it coset E} can be equivalently characterized by:

(i) the integer {\it E} of the form $E=2^{e_0}+2^{e_1}+\cdot \cdot \cdot
+2^{e_{k-1}}.$

(ii) an L-bit string whose 1's are placed at the positions $
\{e_i\}_{i=0,1,...,k-1}$.

(iii) the determinant $A_E$ as defined before.

(iv) the homogeneous linear system (2.1) associated with A$_E,$

\begin{equation}
\left\{ 
\begin{array}{cc}
0= & d_0\alpha ^{t_02^{e_0}}+d_1\alpha ^{t_12^{e_0}}+\cdot \cdot \cdot
+d_{k-1}\alpha ^{t_{k-1}2^{e_0}} \\ 
0= & d_0\alpha ^{t_02^{e_1}}+d_1\alpha ^{t_12^{e_1}}+\cdot \cdot \cdot
+d_{k-1}\alpha ^{t_{k-1}2^{e_1}} \\  
& \vdots \\ 
0= & d_0\alpha ^{t_02^{e_{k-1}}}+d_1\alpha ^{t_12^{e_{k-1}}}+\cdot \cdot
\cdot +d_{k-1}\alpha ^{t_{k-1}2^{e_{k-1}}} 
\end{array}
\right. 
\end{equation}

where $d_j\in GF(2^L)\ \forall j.$

In the sequel these four characterizations will be used indistinctly.
Regarding the use of the binary strings, some additional notation is 
necessary.

Let $E=2^{e_0}+2^{e_1}+\cdot \cdot \cdot +2^{e_{k-1}}$ and $
F=2^{f_0}+2^{f_1}+\cdot \cdot \cdot +2^{f_{l-1}}$ be two L-bit strings of
weight k and l respectively with k$<$l. $E\subset F$ means that $
\{e_i\}_{i=0,1,...,k-1}\subset \{f_i\}_{i=0,1,...,l-1}$. That is, all the
1's in {\it E} are also in {\it F}.

For a set of L-bit strings $\{E_n\}=\{E_1,E_2,...,E_N\}$, $OR[\{E_n\}]$
denotes the L-bit string resulting from a bit wise OR among the L-bit
strings of the set. Obviously, we have that $\forall n\in
\{1,2,...,N\},E_n\subset OR[\{E_n\}].$

Finally, we quote the following definitions and results related to the
global linear complexity of a function with a unique term of maximum order, 
\cite
{Fuster-Caballero}.

A cyclotomic coset is called a {\it fixed-distance coset} if it has an
element{\it \ }$E_d$ of the form $E_d=2^{e_0}+2^{e_1}+\cdot \cdot \cdot
+2^{e_{k-1}},$ with $e_i\equiv d\cdot i\ (mod\ L)\ \forall i\in
\{0,1,...,k-1\}$ and d being a positive integer less than L such that 
(d,L)=1. Its
name is due to the fixed distance d among the positions of the 1's in the
L-bit string representation of $E_d$.

The 1 placed at the position $e_j$ will be called the {\it \ jth-1} 
of the L-bit string associated with the coset $E_d$.

{\bf Theorem 1 }

$f$ is a kth-order function if and only if all the fixed-distance cosets are
nondegenerate.

{\bf Corollary 1}

The global linear complexity $\Lambda $ of the sequence produced by $f$ is
lowerbounded by $\Lambda \geq N_L\cdot L$, where $N_L=\frac{\Phi (L)}2\ $($
\Phi (L)$ being the Euler function). Here $N_L$ represents the number of
fixed-distance cosets and $L$ the number of elements in such cosets.

{\bf Corollary 2}

If $L$ is prime, then the global linear complexity $\Lambda $ of the sequence
generated by $f$ is lowerbounded by $\Lambda \geq {L \choose 2 }$
Remark that these results, which constitute the starting point of the
present work, are independent of the LFSR, the order of f and the particular
form of $f$.

\section{Theoretical Results}

Considering a general function $f$ defined as before, the present work is
concerned with the next simple idea:

Not many degenerate cosets can exist simultaneously.

A proof of this statement can be outlined in three different 
steps. First, the N cosets of a specific
set are supposed to be simultaneously degenerate. Then, it is
proved that only m of these cosets (with m $<$ N) can be
simultaneously degenerate. Consequently, (N-m) cosets contribute to the
global linear complexity of the resulting sequence.

This procedure can be expressed in a more formal way as follows. First of 
all, 
a
new class of cosets is introduced.

Given a fixed-distance coset $E_d=2^{e_0}+2^{e_1}+\cdot \cdot \cdot
+2^{e_{k-1}}$ and $j\in \{0,1,...,k-1\}$, we will call {\it jth-quasi
fixed-distance coset} (for short {\it jth-quasi f-d coset}) to any cyclotomic
coset whose representative element $F_d^j$ is of the form $
F_d^j=2^{f_0}+2^{f_1}+\cdot \cdot \cdot +2^{f_{k-1}}$ such that  $\{e_i\}_{i=0,1,...,k-1}\subset \{f_i\}_{i=0,1,...,k-1}$ ${i\neq j}$. That is, a
jth-quasi f-d coset $F_d^j$ is any cyclotomic coset whose L-bit string
associated contains all the 1's of the L-bit string associated with $E_d$
except for the jth-1.

$\{F_{d,n}^j\}=\{F_{d,1}^j,...,F_{d,N}^j\}$ is used to denote a set of
jth-quasi f-d cosets.

{\bf Lemma 1}

Let $F_d^j$ be any jth-quasi f-d coset, then $A_{F_d^j}$ has at least a
minor of order (k-1) (without the jth-row and an arbitrary ith-column) 
that does not
equal zero:

\begin{equation}
\left| 
\begin{array}{cccccc}
\alpha ^{t_02^{e_0}} & . & \alpha ^{t_{i-1}2^{e_0}} & \alpha
^{t_{i+1}2^{e_0}} & . & \alpha ^{t_{k-1}2^{e_0}} \\ 
. & . & . & . & . & . \\ 
\alpha ^{t_02^{e_{j-1}}} & . & \alpha ^{t_{i-1}2^{e_{j-1}}} & \alpha
^{t_{i+1}2^{e_{j-1}}} & . & \alpha ^{t_{k-1}2^{e_{j-1}}} \\ 
\alpha ^{t_02^{e_{j+1}}} & . & \alpha ^{t_{i-1}2^{e_{j+1}}} & \alpha
^{t_{i+1}2^{e_{j+1}}} & . & \alpha ^{t_{k-1}2^{e_{j+1}}} \\ 
. & . & . & . & . & . \\ 
\alpha ^{t_02^{e_{k-1}}} & . & \alpha ^{t_{i-1}2^{e_{k-1}}} & \alpha
^{t_{i+1}2^{e_{k-1}}} & . & \alpha ^{t_{k-1}2^{e_{k-1}}} 
\end{array}
\right| \neq 0. 
\end{equation}

{\bf Proof }The determinants $A_{F_d^j}$ and $A_{E_d}$ differ exclusively in
the jth-row. Expanding both determinants along the jth-row, we can write $
A_{F_d^j}$ and $A_{E_d}$ in terms of the k minors of order (k-1) of the
form (3.1). The fact that $A_{E_d}\neq 0$ (see Theorem 1) completes the
proof.

The following theorem is the theoretical basis of the LB-algorithm.

{\bf Theorem 2}

Let $E_d$ be any fixed-distance coset and $j\in \{0,1,...,k-1\}$. If for
some set of jth-quasi f-d cosets $\{F_{d,n}^j\}$ there exists at least a
fixed-distance coset $E_{d^{\prime }}$ such that $E_{d^{\prime }}\subset
OR[\{F_{d,n}^j\}],$ then the cosets of $\{F_{d,n}^j\}$ cannot be
simultaneously degenerate.

{\bf Proof }We proceed by contradiction. We assume that the cosets of $
\{F_{d,n}^j\}$ are simultaneously degenerate. This simultaneous degeneration
is equivalent to the existence of a set of homogeneous linear systems
(associated with each determinant $A_{F_{d,n}^j}$) with nontrivial
solutions. All these systems have (k-1) equations in common. Furthermore,
due to Lemma 1, the solutions of each
system are at the same time the joint solutions to all the systems, 
therefore the compatibility of the general system composed of all
the different equations can be easily deduced. Finally, according to the 
starting hypothesis,
the k equations associated with the determinant $A_{d^{\prime }}$ are among
the equations of the general system. This means that a compatible system
has a non-compatible subsystem, which obviously is a contradiction.

The LB-algorithm that is presented in the next section realizes the previous
results by means of the handling of L-bit strings.

\section{The LB-Algorithm}

In this section, the LB-algorithm which computes a lower
bound on the global linear complexity is presented in detail. The
LB-algorithm is based on the previous theorems and corollaries. For every
set of N quasi f-d cosets, the algorithm determines:

(a) the maximum number m of cosets which can be simultaneously degenerate.

(b) the contribution to the global linear complexity of the (N-m) 
remaining cosets which are nondegenerate.

The LB-algorithm converts the linear system (2.1) into an L-bit
string according to the following simple rule: 
the presence of the ith-equation $0=d_0\alpha
^{t_02^{e_i}}+d_1\alpha ^{t_12^{e_i}}+\cdot \cdot \cdot +d_{k-1}\alpha
^{t_{k-1}2^{e_i}}$ in the system implies a 1 in the L-bit
string at the position indicated by $e_i$. Note that, due to the particular
form of
the linear system, squaring the equations 
of the system  (2.1) is equivalent to a left cyclic rotation
in the L-bit string 
associated (Fig. 1). This fact will be used widely throughout the algorithm.

\subsection{Bit Wise Logic Operations}

The LB-algorithm realizes basically three bit wise logic operations AND, OR
and exclusive-OR (denoted by XOR). An interpretation of each operation 
is presented 
in the following.

Given two homogeneous linear systems and their corresponding binary strings,
the AND operation between both strings gives rise to a new homogeneous
linear system whose
equations are common to both systems (Fig. 2 a)).
In the algorithm the logic operation AND will be used to 
check the presence of a particular subsystem inside a general system.

The XOR operation of two L-bit strings associated with both linear systems of
the form (2.1) gives rise to a new system whose equations belong 
exclusively to one of the previous  linear systems (Fig. 2 b)). In the 
following
algorithm the logic operation XOR is used to check if a particular 
coset has been previously studied.

Finally, the OR operation among several L-bit strings 
gives rise to a macrosystem which includes all the equations corresponding
to the systems (Fig. 2 c)). Throughout the algorithm this logic operation
is used as a fundamental tool to check the basic idea of this work:
the simultaneous degeneration of the quasi f-d cosets.

It is clear that the LB-algorithm is based exclusively on the handling of 
L-bit strings 
instead of solving linear systems or computing determinants in a finite field.

\subsection{Notation}

The following notation is used throughout the LB-algorithm.

FDC(i) (i=1,2,...,$N_L$) denotes the L-bit string corresponding to the
ith-fixed-distance coset $E_{d_i}$.

$\Delta $ is a lower bound on the global linear complexity.

MASK(i,j) (j=1,2,...,k-1) denotes the L-bit string obtained from FDC(i) by
replacing the jth-1 by a 0. 
Remark that MASK(i,0) is a shifted version of MASK(i,k-1).

C(i,j) denotes a set of L-bit strings associated with the jth-quasi f-d
cosets $\{F_{d_i,n}^j\}$. Any L-bit string in C(i,j) previously considered
must be eliminated. In order to detect them we operate every L-bit
string in C(i,j) as follows:

1. by means of AND operations with every FDC(i) (i=1,2,...,$N_L$) to discover
the fixed distance cosets

2. by means of XOR operations with every previous MASK. Those cosets that 
produce
a resulting string with a unique 1 must be eliminated from C(i,j) as they 
have been 
already analysed in previous sets  $\{F_{d_i,n}^j\}$.

m is a decreasing counter whose first value (denoted by M) is the number of
L-bit strings in C(i,j) after eliminations.

a(n) (n=1,2,...,${M \choose m}$) denotes each possible M-bit string of weight m.

VOR denotes the string resulting from an OR operation among those m cosets
of C(i,j) indicated by the positions of the 1's in a(n).

VL is a binary variable whose value depends on the AND operation between VOR
and each FDC(i).

\subsection{Algorithm}

The LB-algorithm INPUTS are L (LFSR's length) and k (order of the function)
with 2$<$ k$<$ L-2, and its OUTPUT is the lower bound of
the global linear complexity $\Delta $.

Fig. 3 shows the LB-algorithm whose Steps 1 and 2 can be described as
follows.

{\bf Step 1}

Compute the $N_L$ values of d.

Generate the FDC(i) (i=1,2,...,$N_L$).

Initialize the lower bound $\Delta =L\cdot N_L$.

{\bf Step 2}

Generate MASK(i,j) (i=1,2,...,$N_L$;\ j=1,2,...,k-1).

Initialize the counter m=L-k.

Generate the set C(i,j).

Realize the AND between every FDC(l) (l=1,2,...,$N_L$) and every coset of
C(i,j). If any result equals FDC(l), then the corresponding coset in C(i,j)
is eliminated and m=m-1.

Realize the XOR between every MASK(o,p) (o=1,2,...,i-1,\ p=1,2,...,k-1;\ 
o=i,\
p=1,2,...,j-1) and every coset of C(i,j). If any result 
has a unique 1, then the corresponding coset in C(i,j) is
eliminated and m=m-1.

\subsection{Example}

Fig.  4 shows the results obtained from the LB-algorithm for L=11 and k=6.

Since the LB-algorithm is independent of the specific function and minimal
polynomial of the LFSR, the lower bound obtained is valid for any arbitrary
nonlinear
function with a unique term of maximum order 6 and for any maximal-length
LFSR of length 11.

If we had used the root presence test to obtain the same result, we would
have had to compute (for each function of order 6 and each maximal-length
LFSR of length 11) at least 22 determinants of order 6 in $GF(2^{11})$. This
would have implied more than a million arithmetic operations in a finite
field, \cite{Knuth}. According to the present algorithm, the numerical result
obtained is independent of the function and the maximal-length LFSR.

\subsection{Discussion}

The main facts concerning the performance of the algorithm are summarized in
this section.

The LB-algorithm is divided into two stages. The first stage includes the
generation and `debugger' of the cosets to be analysed. The second stage is
concerned with the simultaneous degenerations of the different sets of
cosets. In the second stage a `sweep' of some sets of cosets is carried out,
which permits their use later on the algorithm.

Regarding the required memory, note that only the L-bit strings 
MASK(i,j) (but not
the cosets C(i,j)) have to be stored. This means keeping one out of (L-k)
cosets analysed.

In order to handle the cosets of C(i,j), the more suitable structure of
information is a list. This structure seems also adequate to select, through
the codification a(n), the cosets involved in each OR operation. On the
other hand, in order to generate the successive strings a(n), backtracking 
can be
used.

It can also be determined that the LB-algorithm has a maximum
computational complexity of order $O(2^{L-k})$, where L denotes the length
of the LFSR and k is the order of the function. In order to estimate this
value, it has been assumed the `worst possible case', which involves a
number of logic operations given by $N_L(k-1)[{M \choose M}+ {M \choose {M-1}}+\
\cdot \cdot \cdot +\  {M \choose 2}]=\ N_L(k-1)(2^M-M)\leq \ N_L(k-1)2^{L-k}$.
However, from the experimental results it can be deduced that the running
time of the LB-algorithm depends on the real number of bit wise
operations among the different L-bit strings, which is much less.
As an illustrative example we can say that for L=53 and k=27 the number of 
logic operations is only
$N_{53}(27-1)[ {25 \choose 25}+ {25 \choose 24}+
 {25 \choose 23}+ {25 \choose 22}+{25 \choose 21}]\leq 
\ N_{53} \cdot 26 \cdot 2^{26}$.

Furthermore the following three considerations must be taken into account.
First, for each pair of values (L,k), the LB-algorithm has to be used only
once. Second, it will be used only with relatively small inputs. And third,
a high bound obtained for specific values of L and k will encourage the
designer of running-key generators to use nonlinear filter with a unique
term of maximum order k applied to any maximal-length LFSR of length L.

The LB-algorithm has been implemented on a DEC work-station and several
experiments over values of L primes have been carried out to evaluate it.
The effect of this choice is twofold. On the one hand, it simplifies the
computation of the $N_L$ values of d in Step 1, and on the other hand, the
more fixed-distance cosets there are the higher bounds the algorithm
computes.

The following table shows some experimental results.

\begin{center}
\begin{tabular}{|c|cccccccc|}
\hline
L & 11 & 17 & 23 & 29 & 37 & 43 & 47 & 53   \\ 
\hline
k & 6 & 9 & 12 & 15 & 19 & 22 & 24 & 27 \\
\hline 
Bound & 242 & 3128 & 8349 & 22330 & 47952 & 75852 & 99405 & 143206 \\
\hline
\end{tabular}

Table 1: Lower bounds on the global linear complexity     

\end{center}

According to the values shown, the LB-algorithm is believed to be
quite efficient to lowerbound the global
linear complexity of the filtered sequences. The growth of the bound
observed can be approximated by the curve of Fig 5, which has been obtained
through regression analysis for the linear model. This approximation let us
estimate a bound above 500000 for L=89.

In conclusion, the main result deduced from the LB-algorithm is reliability 
for the
nonlinear filter. Thanks to it a designer of nonlinear filter generators
could carry out the following steps:

1.- Find values of L and k that produce a high lower bound,

2.- Choose any nonlinear function of a smaller order than k,

3.- Add it to any kth-order product and

4.- Apply the resulting nonlinear function to any maximal-length LFSR of
length L.

In this way the designer would obtain a sequence with a guaranteed 
large global linear
complexity.

\section{Conclusions}

Our research has highlighted the problem of the global linear complexity of
the nonlinear filter generators. In addition, a new algorithm, the so-called
LB-algorithm, to lowerbound the global linear complexity has
been presented.

This proposal differs from existing schemes in different aspects. Firstly,
unlike the well-known Berlekamp-Massey's algorithm \cite{Massey}, we do not
consider the digits of the output sequence but the characteristics of the
nonlinear filter. Secondly, the proposed algorithm indeed does not require
any condition on the LFSR's stages involved, as do \cite{Kumar-Scholtz} and 
\cite{Massey-Serconek}. Therefore the obtained bounds are valid for any
nonlinear function with a unique term of maximum order. Finally, this
work is based on the handling of L-bit strings instead of computing
determinants in a finite field (Rueppel's method, \cite{Rueppel}), 
which seems to be much more adequate for software simulation
and/or hardware implementation.

Large lower bounds for the global linear complexity have been obtained from
the LB-algorithm without imposing any restriction on the function or the
polynomial. This fact ensures the reliability of the nonlinear state-filter
generators for cryptographic application.

This investigation has left as open problem the study of the remaining
cosets that the LB-algorithm does not analyse.

\end{document}